\documentclass[aps,prl,twocolumn,showpacs]{revtex4-1}

\usepackage{units}
\usepackage{xspace}
\usepackage{graphicx}
\usepackage{mathrsfs}
\usepackage{subfigure}
\usepackage[hyperindex,breaklinks]{hyperref} 

\usepackage{preprintcover} 
\PreprintCoverPaperTitle{Search for supersymmetry in events with three leptons and missing transverse momentum in $\sqrt{s}\,$$=\,$7$\,$TeV $pp$ collisions \\ with the ATLAS detector}  
\PreprintIdNumber{CERN-PH-EP-2012-098}  
\PreprintCoverAbstract{A search for the weak production of charginos and neutralinos decaying to a final state with three leptons (electrons or muons) and missing transverse momentum is presented. The analysis uses 2.06$\,$fb$^{-1}$ of 
 $\sqrt{s}\,$$=\,$7$\,$TeV proton-proton collision data delivered by the Large Hadron Collider and recorded with the ATLAS detector. 
Observations are consistent with  standard model expectations in two signal regions that are either depleted or enriched in $Z$-boson decays. 
Upper limits at 95\% confidence level are set in R-parity conserving phenomenological minimal supersymmetric and  
simplified models. For the simplified models, degenerate lightest chargino and next-to-lightest neutralino masses 
up to 300$\,$GeV are excluded for mass differences from the lightest neutralino up to 300$\,$GeV.} 
\PreprintJournalName{Physical Review Letters}  

\newcommand{\Alpgen}{{\tt ALPGEN}\xspace}
\newcommand{\Jimmy}{{\tt JIMMY}\xspace}
\newcommand{\Mcatnlo}{{\tt MC@NLO}\xspace}
\newcommand{\Herwig}{{\tt HERWIG}\xspace}
\newcommand{\Herwigpp}{{\tt HERWIG++}\xspace}

\newcommand{\Prospino}{{\tt PROSPINO}\xspace}
\newcommand{\Madgraph}{{\tt MadGraph}\xspace}
\newcommand{\MRST}{{\tt MRST 2007 LO$^*$}\xspace}
\newcommand{\CTEQSixL}{{\tt CTEQ6L1}\xspace}
\newcommand{\CTEQSix}{{\tt CTEQ6.6}\xspace}
\newcommand{\PYTHIA}{{\tt PYTHIA}\xspace}
\newcommand{\GEANT}{{\tt GEANT4}\xspace}
\newcommand{\MSTW}{{\tt MSTW}\xspace}
\newcommand{\MCFM}{{\tt MCFM}\xspace}
\newcommand{\fewz}{{\tt FEWZ}\xspace}
\newcommand{\antibar}[1]{\ensuremath{#1\bar{#1}}}
\newcommand{\ttbar}{\antibar{t}\xspace}

\newcommand{\ninoone}{\ensuremath{\tilde{\chi}_1^0}\xspace}
\newcommand{\ninotwo}{\ensuremath{\tilde{\chi}_2^0}\xspace}
\newcommand{\chinoone}{\ensuremath{\tilde{\chi}_1^\pm}\xspace}

\newcommand{\met}{\ensuremath{E_{\mathrm{T}}^{\mathrm{miss}}}\xspace}

\def\ninoi{\ensuremath{\mathchoice
      {\displaystyle\raise.4ex\hbox{$\displaystyle\tilde\chi^0_j$}}%
         {\textstyle\raise.4ex\hbox{$\textstyle\tilde\chi^0_j$}}%
       {\scriptstyle\raise.3ex\hbox{$\scriptstyle\tilde\chi^0_j$}}%
 {\scriptscriptstyle\raise.3ex\hbox{$\scriptscriptstyle\tilde\chi^0_j$}}}}

\def\chinojpm{\ensuremath{\mathchoice%
      {\displaystyle\raise.4ex\hbox{$\displaystyle\tilde\chi^\pm_i$}}%
         {\textstyle\raise.4ex\hbox{$\textstyle\tilde\chi^\pm_i$}}%
       {\scriptstyle\raise.3ex\hbox{$\scriptstyle\tilde\chi^\pm_i$}}%
{\scriptscriptstyle\raise.3ex\hbox{$\scriptscriptstyle\tilde\chi^\pm_1$}}}}

\begin{document}
 \title{Search for supersymmetry in events with three leptons and missing transverse momentum in $\sqrt{s}\,$$=\,$7$\,$TeV $pp$ collisions with the ATLAS detector}
\author{The ATLAS Collaboration}
\begin{abstract}
A search
for the weak production of charginos and neutralinos decaying to a final state with three leptons (electrons or muons) and missing transverse momentum is presented. The analysis uses 2.06$\,$fb$^{-1}$ of 
 $\sqrt{s}\,$$=\,$7$\,$TeV proton-proton collision data delivered by the Large Hadron Collider and recorded with the ATLAS detector. 
Observations are consistent with  standard model expectations in two signal regions that are either depleted or enriched in $Z$-boson decays. 
Upper limits at 95\% confidence level are set in R-parity conserving phenomenological minimal supersymmetric and  
simplified models. For the simplified models, degenerate lightest chargino and next-to-lightest neutralino masses 
up to 300$\,$GeV are excluded for mass differences from the lightest neutralino up to 300$\,$GeV.
\end{abstract}
\maketitle
Supersymmetric (SUSY)
extensions~\cite{Miyazawa:1966,Ramond:1971gb,Golfand:1971iw,Neveu:1971rx,Neveu:1971iv,Gervais:1971ji,Volkov:1973ix,Wess:1973kz,Wess:1974tw} 
of the standard model (SM) naturally address the gauge hierarchy problem~\cite{hierarchy1,hierarchy2,hierarchy3} 
  by postulating the existence of  SUSY particles, or ``sparticles",  with  spin differing by one-half unit with respect to that of their SM partner. If 
R-parity~\cite{Fayet:1976et,Fayet:1977yc,Farrar:1978xj,Fayet:1979sa,Dimopoulos:1981zb} is conserved, sparticles can only be pair-produced and will eventually decay into SM particles and the lightest 
SUSY particle (LSP) which is stable.  
Charginos ($\chinojpm$, $i=1,2$) and neutralinos ($\ninoi$, $j=1,2,3,4$) are the mass eigenstates formed from the linear superposition of the SUSY partners of the Higgs and  electroweak gauge bosons. These are the higgsinos, and the winos, zino, and bino,  collectively known as  gauginos.  
Naturalness  requires $\chinojpm$ and $\ninoi$ (and third generation sparticles) to have masses in the hundreds of GeV range~\cite{naturalness}. 
In scenarios where first and second generation sfermion masses are larger than few TeVs, the direct production of weak gauginos may be the dominant SUSY process at the Large Hadron Collider (LHC). 

Leptonic decays of charginos include sneutrinos ($\tilde{\nu}\ell^{\pm}$), sleptons ($\tilde{\ell}^{\pm}\nu$)   
or $W$ bosons ($W^{\pm (*)}\ninoone$), while those of unstable neutralinos include sleptons  ($\ell\tilde{\ell^{\pm}}$) 
or $Z$ bosons ($Z^{(*)}\,\ninoone$). When both gauginos decay leptonically, a distinctive signature with three leptons and significant missing 
transverse momentum can be observed, the latter originating from the two undetected LSPs and the neutrinos.

This letter presents the first  search with the ATLAS detector 
for the weak production of charginos and neutralinos decaying to a final state with three leptons (electrons or muons) and missing transverse momentum. 
The analysis is based on 2.06$\,$fb$^{-1}$ of proton-proton collision data delivered by the LHC at a 
center-of-mass energy $\sqrt{s}\,$$=\,$7$\,$TeV between March and August 2011.  The search 
 significantly 
extends the current mass limits on charginos and neutralinos~\cite{lep,D0-2009, CDF-2008,ATLAS2lep2011} and yields sensitivity in the mass region preferred by naturalness.

In this analysis, observations are interpreted in the 
phenomenological minimal supersymmetric SM (pMSSM \cite{pMSSM}) and in simplified models~\cite{simplModels1}.  
In the pMSSM the masses of the $\chinojpm$ and $\ninoi$ depend on the gaugino masses $M_1$ and $M_2$, the Higgs mass parameter $|\mu |$, 
and $\tan\beta$, the ratio of the expectation 
values of the two Higgs doublets. 
The masses of the gluinos, squarks and left-handed sleptons are chosen to be larger than 1$\,$TeV, 
while the  right-handed sleptons (including third-generation ones) are assumed to be degenerate with $m_{\tilde{\ell}_R}\,$$=\,$$(m_{\tilde{\chi}_2^{0}} + m_{\tilde{\chi}_1^{0}} )/2$. In these scenarios, decays to sleptons are favored. 
In the simplified models, the masses of the relevant particles (mass degenerate wino-like $\chinoone$ and $\ninotwo$; bino-like $\ninoone$; $\tilde{\nu}$; $\tilde{\ell}_L$) are the only 
free parameters 
of 
the theory. The $\chinoone$ and $\ninotwo$ are produced  
via the $s$-channel exchange of a virtual
gauge boson and  decay via  left-handed sleptons, including staus, and sneutrinos of mass  $m_{\tilde{\nu}}=m_{\tilde{\ell}_L}\,$$=\,$$(m_{\tilde{\chi}_1^{0}} + m_{\tilde{\chi}_1^{\pm}} )/2$) with a  
branching ratio of 50\%  each. 

ATLAS~\cite{atlas-det} is a multipurpose particle detector with forward-backward symmetric cylindrical geometry. It includes  an inner tracker (ID) immersed in a 2$\,$T magnetic field  providing precision tracking of charged particles for pseudorapidities $|\eta|\,$$<\,$2.5~\cite{coordSys}.
 Calorimeter systems with either liquid argon or scintillating tiles as the active media provide energy measurements over the range $|\eta|\,$$<\,$4.9. 
The muon detectors outside the calorimeters are contained in a  toroidal magnetic field   produced by air-core superconducting magnets with field integrals varying from 1 to 8$\,$T$\cdot$m. They provide trigger and high-precision tracking 
capabilities for $|\eta|\,$$<\,$2.4 and $|\eta|\,$$<\,$2.7, respectively.
Electrons must satisfy tight identification criteria and fulfil $|\eta|\,$$<\,$2.47 and $E_{\rm T}\,$$>\,$10$\,$GeV, where 
$|\eta|$ and $E_{\rm T}$ are determined from the calibrated clustered energy deposits in the electromagnetic calorimeter matched to an ID track.
Muons are reconstructed by combining tracks in the ID and tracks in the muon spectrometer~\cite{muon}. Reconstructed muons are considered as candidates if they have  transverse momentum $p_{\rm T}\,$$>\,$10$\,$GeV, $|\eta|\,$$<\,$2.4, and transverse impact parameter with respect to the primary vertex $|d_0|\,$$<\,$0.2$\,$mm. 
``Tagged" leptons are electrons and muons, well separated from each other and from candidate jets. ``Signal'' leptons are tagged leptons for which the scalar sum of the tracks' transverse momenta within  $\Delta R\,$$\equiv\,$$\sqrt{(\Delta\phi)^2+(\Delta\eta)^2}\,$$<\,$0.2
around the lepton candidate is 
less than 10\% of the $E_{\rm T}$ for 
electrons, and less than 1.8$\,$GeV for muons. 
Jets are reconstructed from  clustered energy deposits calibrated at the electromagnetic scale  using the anti-$k_t$ algorithm~\cite{Cacciari:2008gp} with a radius parameter of 0.4. 
The jet energy is corrected to account for the non-compensating nature of the calorimeter using correction factors 
obtained from Monte Carlo (MC) simulation and parameterized as a function of the jet $E_{\rm T}$ and $\eta$~\cite{fakeCleanUp}.
Jets considered in this analysis have $E_{\rm T}\,$$>\,$20$\,$GeV and $|\eta|\,$$<\,$2.8. 
Jets are identified as containing a $b$-quark, and thus called ``$b$-tagged'', using
a multivariate technique based on quantities
such as the impact parameter of the tracks associated to the secondary vertex,
tracks in jet and other jet shape information, consistent with the expected
topology of $b$-quark decays. 
The $b$-tagging algorithm~\cite{btag}  correctly identifies $b$-quark jets in top decays with an efficiency of 60\%
and misidentifies jets containing light flavor quarks and gluons with a rate of $<\,$1\%, for jets with $|\eta|\,$$<\,$2.5 and jet $E_{\rm T}\,$$>\,$20$\,$GeV. 
The missing transverse momentum, $\met$, is the magnitude of the vector sum of the transverse momentum or transverse energy 
of all $p_{\rm T}\,$$>\,$10$\,$GeV muons, $E_{\rm T}\,$$>\,$10$\,$GeV electrons, 
 $E_{\rm T}\,$$>\,$20$\,$GeV jets, and calibrated 
calorimeter clusters with $|\eta|\,$$<\,$4.5 not associated to these objects~\cite{metPerf}. 

Several MC generators are used to simulate SM processes and SUSY signals relevant for this analysis.
\Herwig~\cite{herwig} is used to simulate diboson processes ($WW^{(*)}, ZZ^{(*)}, WZ^{(*)}$), while \Madgraph~\cite{madgraph} is used for the $t\bar{t}W^{(*)}/Z^{(*)}$ processes.
\Mcatnlo~\cite{mcatnlo} is chosen for the simulation of single and pair production of top quarks, while \Alpgen~\cite{alpgen} is used to simulate $W^{(*)}/Z^{(*)}$+jets. 
Expected diboson yields are normalized using 
next-to-leading order (NLO) QCD predictions obtained with \MCFM~\cite{mcfm1,mcfm2}. The top-quark pair-production contribution is normalized to  approximate 
next-to-next-to-leading (NNLO) order calculations~\cite{hathor} and the $t\bar{t}W^{(*)}/Z^{(*)}$ contributions are normalized to NLO results~\cite{ttV}. The QCD NNLO \fewz~\cite{fewz1,fewz2} and \MCFM 
cross-sections  
are used for 
NNLO normalization of the $Z$+light-flavor jets and $Z$+heavy-flavor jets processes, respectively.
The choice of the parton distribution functions (PDFs) depends on the generator. \MRST~\cite{MRST2007} sets are used for \Herwig, \CTEQSixL~\cite{CTEQ6L1} with \Madgraph\ and \Alpgen, and 
\CTEQSix~\cite{CTEQ66} with \Mcatnlo. 
The pMSSM and  simplified model samples are produced with \Herwig\ and \Herwigpp~\cite{herwigplusplus}, respectively, and the yields are normalized to the NLO cross-sections obtained from 
\Prospino~\cite{prospino} using the PDF set \CTEQSix with  the renormalization/factorization scales 
set to the average of the relevant gaugino 
masses. 
Fragmentation and hadronization for the \Alpgen and \Mcatnlo  (\Madgraph) samples are performed with \Herwig  (\PYTHIA~\cite{pythia}). For all MC samples, the propagation of particles through the 
ATLAS detector is modeled 
using \GEANT~\cite{Agostinelli:2002hh,simulation}. The effect of multiple proton-proton collisions from the same or different bunch crossings is incorporated into the simulation by overlaying additional minimum bias 
 events onto hard scatter events using \Jimmy~\cite{jimmy}. Simulated events are weighted
 to match the distribution of the  mean number of interactions per bunch crossing observed in data.

The data sample was collected with a single-muon trigger ($p_{\rm T}$$>$18$\,$GeV) or a  single-electron trigger ($E_{\rm T}$$>$20 or 22$\,$GeV, depending on the instantaneous luminosity). 
At least one signal lepton is required to have triggered the
event 
and have $p_{\rm T}$ ($E_{\rm T}$) above 20$\,$GeV (25$\,$GeV) for muons (electrons). 
Events recorded during 
normal running conditions are analyzed if at least one of the reconstructed primary vertices  
 has more than four tracks associated  
to it. Events containing jets with $|\eta|<$$\,$4.9  and failing the quality criteria described in Ref.~\cite{fakeCleanUp} are rejected to suppress both collisional and non-collisional background. 
Selected events must contain exactly three signal leptons.
As leptonic decays of $\ninoi$ yield same-flavor opposite-sign (SFOS) lepton pairs, the presence of at least one such pair is required. The invariant mass of any SFOS lepton pair must be 
above 20$\,$~GeV to suppress  background from low mass resonances 
and the missing transverse momentum must satisfy $\met\,$$>\,$50$\,$GeV.

Two signal regions are then defined: a ``$Z$-depleted'' region (SR1), with no SFOS pairs having invariant mass within 10$\,$GeV of the nominal $Z$-boson mass; and a ``$Z$-enriched'' one (SR2), where at 
least one SFOS pair has an invariant mass within 10$\,$GeV of the  $Z$-boson mass. Events in SR1 are further required to contain no $b$-tagged jets to suppress contributions from $b$-jet-rich 
backgrounds, where a fake lepton could originate from a heavy-flavor decay. 
The SR1  and SR2 selections target SUSY events with intermediate slepton or on-mass-shell $Z$-boson decays, respectively. 

Several SM processes contribute to the background in SR1 and SR2. A background process is considered ``irreducible'' if it leads to events with three real and isolated leptons, referred to as  ``real'' leptons below.  These include diboson ($WZ^{(*)}$ and $ZZ^{(*)}$) and $t\bar{t}W/Z^{(*)}$ production, where the gauge boson may be produced off-mass-shell. Their contribution is determined using the corresponding MC samples  
, for which lepton and jet selection efficiencies are corrected to account for differences with respect to data.  
   A ``reducible" process has at least one ``fake" object, 
that is either a lepton from a semileptonic decay of a heavy-flavor quark or an electron from an isolated photon conversion.  The contribution from misidentified light-flavor quarks   
 is negligible. The reducible background includes single- and pair-production of top-quark and $WW^{(*)}$ or $W^{(*)}/Z^{(*)}$ produced in association with jets or photons. The dominant component is the production of top quarks, with a contribution of 1\% or less from $Z^{(*)}$+jets production. The reducible background is estimated using a ``matrix method'' similar to that described in Ref.~\cite{mm}. 

In this implementation of the matrix method, the signal lepton with the highest $p_{\rm T}$ or $E_{\rm T}$ is taken to be real, which is a valid assumption in 99\% of the cases, based on MC studies. 
The number of observed events with one or two fakes is then extracted from a system of linear equations linking
the number of events with two additional signal or tagged candidates to the number of events
with two additional candidates that are either real or fake. 
The coefficients of the linear equations are functions of the real lepton identification efficiencies and of the fake object misidentification probabilities. The identification efficiency is measured in data using lepton candidates from 
$Z\rightarrow\ell\ell$ decays.

Misidentification probabilities for each fake  type (heavy flavor, conversion) and for each reducible background process are  obtained using  simulated events with one signal and two tagged leptons. These misidentification probabilities are then corrected using the ratio (fake scale factor)  of the misidentification probability in data and that in MC simulation obtained in dedicated control samples. For heavy flavor fakes, the correction factor is measured in $b\bar{b}$ events, while for conversion fakes it is determined in a sample of photons radiated from a muon in  $Z\rightarrow\mu\mu$ decays. A weighted average misidentication probability is then calculated by 
weighting the corrected type- and process-dependent misidentification probabilities according to the process cross-section.

An additional source of background is due to events with two signal leptons and one virtual photon converting into two muons, 
one with $p_{\rm T}$ above 10$\,$GeV. The contribution from events in which both muons from the virtual photons have $p_{\rm T}$ above 10$\,$GeV is negligible due to the requirement on the dilepton pair invariant mass. For events with only one muon above threshold, an upper limit of $0.5\pm 0.5$ in SR1 and of $0.7\pm 0.7$ in SR2 is obtained from
data as follows.
The 
number 
of observed events with exactly two signal leptons and $\met\,$$>\,$50$\,$GeV is rescaled by the probability that any of the signal leptons could have radiated the converted photon. 
This probability is measured in events with $\met\,$$<\,$50$\,$GeV as the ratio of number of events with three signal muons 
with trilepton invariant mass within 10$\,$GeV of the nominal $Z$ boson mass to the 
number of events with two signal muons having the dilepton mass in the same mass window. 

The matrix method has been tested using MC events and shown to be accurate within 2\%.
The background predictions have been validated in a  region dominated by $Z^{(*)}$+jets production (VR1: three signal leptons, 30$<\,$$\met\,$$<\,$50$\,$GeV) and in one dominated by top pair-production (VR2: three signal leptons, SFOS lepton pairs vetoed, $\met\,$$>\,$50$\,$GeV). The data and predictions are in agreement within the quoted statistical and systematic uncertainties as shown in Table~\ref{tab:SRs}.

Several sources of systematic uncertainty are considered in the signal regions. 
The systematic uncertainties affecting the MC based estimates (irreducible background yield, misidentification probabilities, signal yield) include the  theoretical cross section uncertainty due to scale and PDFs,  the acceptance uncertainty due to PDFs,  jet energy scale,  jet energy resolution, lepton energy scale, lepton energy resolution, lepton efficiency, $b$-tagging efficiency, event quality selection, and the uncertainty  on the  luminosity. 
In SR1, the total uncertainty on the irreducible background is 17\%. This includes the uncertainty on the acceptance due to PDFs (14\%),  that on the theoretical cross section due to scale and PDFs (7\%), and that from the limited number of simulated events (10\%), while all the remaining uncertainties on the irreducible background in this signal region range between 0.5-5\%. 
The total uncertainty on the reducible background is 29\%. This includes an uncertainty on the object misidentification probability of 10-30\% from the sources listed above. 
The uncertainty from the dependence of the misidentification probability on
$\met$  (0.4--17\%) and the uncertainty on the fake scale factors (10--50\%) are also included in the total, together with the uncertainty from 
the limited number of data events with three tagged leptons, of which at least one is a signal lepton.
The total uncertainties on the signal cross-section range between 10-15\%. These include uncertainties due to the renormalization and factorization scale, $\alpha_S$, and PDFs. The maximum uncertainty obtained from either the \CTEQSix\ or the \MSTW~\cite{MSTW}  PDF set is used. 
In SR2, the values of systematic uncertainties  are similar to those obtained in SR1. The only exceptions are the uncertainty from the limited number of  simulated events (4\%) and the uncertainty on the reducible background (52\%). In all of the above, the value used for the uncertainty on the luminosity is 3.7\%.

\begin{figure}[h]
 \includegraphics[width=9cm]{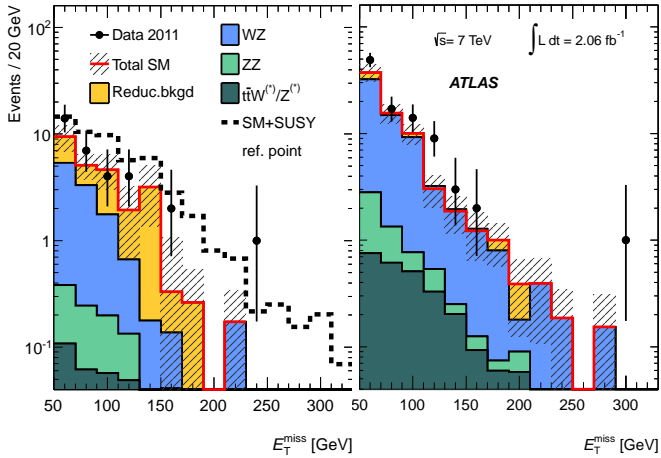}
\caption{\label{fig:met} $\met$ distributions for events in signal regions SR1 (left) and SR2 (right). 
The error band  includes both statistical and systematic uncertainty, while the errors on the  data points are statistical only.
The SUSY reference point used in SR1 is described in the text. 
}
\end{figure}

The numbers of observed events and the prediction for SM backgrounds in SR1 and SR2 are reported in Table~\ref{tab:SRs}. The probability that the background fluctuates to the observed number of events or higher is calculated in the frequentist approach and found to be 19\% in SR1 and 6\% in SR2. The distributions of the $\met$ in the two signal regions are presented in Fig.~\ref{fig:met}.  The 
yield in SR1 for one of the simplified model scenarios ($m_{\chinoone}, m_{\ninotwo} , m_{\tilde{\ell}_L}, m_{\ninoone}\,$$=\,$250, 250, 175, 100$\,$GeV) is also shown for illustration 
purposes.
\begin{table}[h]
\begin{center}
 \caption{Expected numbers of events from SM backgrounds (Bkg.) and observed numbers of events in data, for 2.06~fb$^{-1}$,
 in control regions VR1 and VR2, and in signal regions SR1 and SR2. Both statistical and systematic uncertainties are included. \label{tab:SRs}}
 \begin{tabular}{l r@{$\pm$}l r@{$\pm$}lr@{$\pm$}l r@{$\pm$}l}
   \hline\hline
    Selection		&\multicolumn{2}{c}{VR1}&\multicolumn{2}{c}{VR2}&\multicolumn{2}{c}{SR1}&\multicolumn{2}{c}{SR2} \\ 
    \hline
    $\ttbar W^{(*)}/Z^{(*)}$		& 1.4 & 1.1		& 0.7 & 0.6		& 0.4 & 0.3 	& 2.7 & 2.1 \\ 
    $ZZ^{(*)}$		& 6.7 & 1.5		& 0.03 & 0.04		& 0.7 & 0.2		& 3.4 & 0.8 \\
    $WZ^{(*)}$		& 61& 11		& 0.4 & 0.2		& 11 & 2		& 58 & 11 \\ 
    Reducible Bkg.	& 56 & 35		& 14 & 9		& 14& 4		& 7.5 & 3.9 \\
    Total Bkg.		& 125 & 37		& 15 & 9		& 26 & 5		& 72 & 12 \\
    Data		&\multicolumn{2}{c}{122}&\multicolumn{2}{c}{12}	&\multicolumn{2}{c}{32}	&\multicolumn{2}{c}{95} \\
   \hline\hline
 \end{tabular}
 \end{center}
\end{table}

\begin{figure}[h]
  \centering\includegraphics[width=8cm]{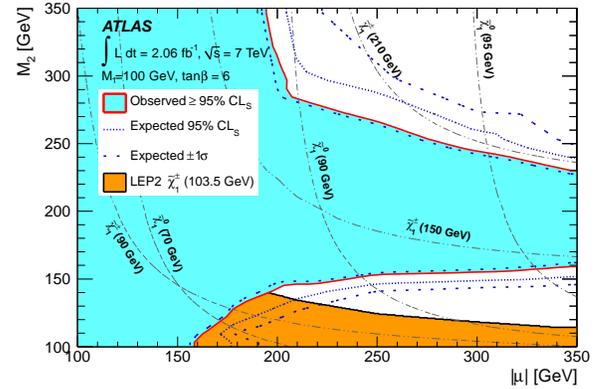}  \centering\includegraphics[width=8cm]{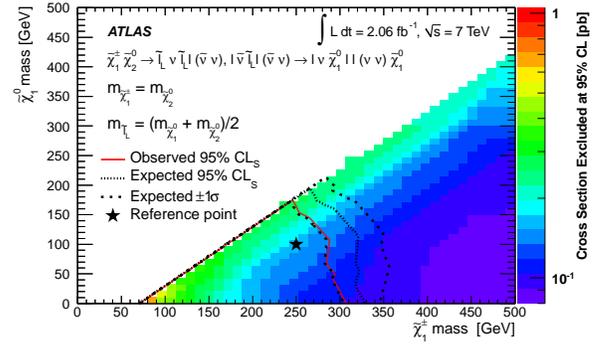}
  \caption{Observed and expected 95\% CL limit contours for chargino and neutralino production in the pMSSM (upper) and simplified model (lower) scenarios. For the simplified models, 
the 95\% CL upper limit on the production cross-section is also shown. Interpolation is used to account for the discreteness of the signal grids. \label{fig:limits}}
\end{figure}

No significant excess of events is found in either signal region. 
Upper limits on the visible production cross-section of 9.9$\,$fb in SR1 and 23.8$\,$fb in SR2 are placed 
at 95\% confidence level  (CL) with the modified frequentist CL$_s$ 
prescription~\cite{CLs}. 
 No corrections for the effects of experimental resolution,
acceptance and efficiency are applied. All  systematic uncertainties and their correlations are taken into account via nuisance parameters.
 The corresponding expected limits are 7.1$\,$fb and  14.1$\,$fb, respectively. In SR2, the observed limit on the visible cross-section is less stringent than the expected limit because of the upwards fluctuation in the number of events in data with respect to the expected background. 
SR1 provides better sensitivity in the  parameter space considered and the limits  are interpreted in  simplified models and 
 pMSSM scenarios with $M_1$$=\,$100$\,$GeV and $\tan\beta$\,=\,6 (Fig.~\ref{fig:limits}). 
 The chosen $M_1$ value leads to a sizable mass splitting between  \chinoone and \ninoone and 
 therefore to a large acceptance.  The value of $\tan\beta$  does not have a significant impact on 
  $\sigma(pp\rightarrow\chinojpm\ninoi)\times BR(\chinojpm\ninoi\rightarrow\ell\ell\ell\ninoone)$, which varies by $\sim$10\% if $\tan\beta$ is raised to 10.  

In the simplified models, 
degenerate  \chinoone\  and  \ninotwo\ masses 
up to 300$\,$GeV are excluded for large mass differences from the \ninoone.  
Care has to be taken when interpreting the simplified model limit in the context of a pMSSM scenario, where 
the mass of the sneutrino is lighter than the mass of the left-handed slepton, leading to higher lepton momenta from chargino decays and to a change in the 
branching ratios of the \ninotwo.

In summary,
results from the first ATLAS search 
for the weak production of charginos and neutralinos decaying to a final state with three leptons (electrons or muons) and missing transverse momentum are reported. The analysis is based on  
2.06$\,$fb$^{-1}$ of  proton-proton collision data delivered by the LHC at  $\sqrt{s}=$7$\,$TeV.  
No significant excess of events is found in data, where upwards fluctuations of less than 2-sigma are observed.  The null result is interpreted in pMSSM and in simplified models. For the simplified models with intermediate sleptons considered in this paper, degenerate lightest chargino and next-to-lightest neutralino masses
are excluded up to 300$\,$GeV for mass differences to the lightest neutralino up to 300$\,$GeV.

\section{Acknowledgements}

We thank CERN for the very successful operation of the LHC, as well as the
support staff from our institutions without whom ATLAS could not be
operated efficiently.

We acknowledge the support of ANPCyT, Argentina; YerPhI, Armenia; ARC,
Australia; BMWF, Austria; ANAS, Azerbaijan; SSTC, Belarus; CNPq and FAPESP,
Brazil; NSERC, NRC and CFI, Canada; CERN; CONICYT, Chile; CAS, MOST and NSFC,
China; COLCIENCIAS, Colombia; MSMT CR, MPO CR and VSC CR, Czech Republic;
DNRF, DNSRC and Lundbeck Foundation, Denmark; EPLANET and ERC, European Union;
IN2P3-CNRS, CEA-DSM/IRFU, France; GNAS, Georgia; BMBF, DFG, HGF, MPG and AvH
Foundation, Germany; GSRT, Greece; ISF, MINERVA, GIF, DIP and Benoziyo Center,
Israel; INFN, Italy; MEXT and JSPS, Japan; CNRST, Morocco; FOM and NWO,
Netherlands; RCN, Norway; MNiSW, Poland; GRICES and FCT, Portugal; MERYS
(MECTS), Romania; MES of Russia and ROSATOM, Russian Federation; JINR; MSTD,
Serbia; MSSR, Slovakia; ARRS and MVZT, Slovenia; DST/NRF, South Africa;
MICINN, Spain; SRC and Wallenberg Foundation, Sweden; SER, SNSF and Cantons of
Bern and Geneva, Switzerland; NSC, Taiwan; TAEK, Turkey; STFC, the Royal
Society and Leverhulme Trust, United Kingdom; DOE and NSF, United States of
America.

The crucial computing support from all WLCG partners is acknowledged
gratefully, in particular from CERN and the ATLAS Tier-1 facilities at
TRIUMF (Canada), NDGF (Denmark, Norway, Sweden), CC-IN2P3 (France),
KIT/GridKA (Germany), INFN-CNAF (Italy), NL-T1 (Netherlands), PIC (Spain),
ASGC (Taiwan), RAL (UK) and BNL (USA) and in the Tier-2 facilities
worldwide.

\bibliographystyle{apsrev4-1}
\bibliography{atlas_susy_trilepton}

\begin{thebibliography}{10}
\bibitem{coordSys}%
  \BibitemOpen
  \bibinfo {howpublished} {ATLAS uses a right-handed coordinate system with its
  origin at the nominal interaction point (IP) in the centre of the detector
  and the $z$-axis along the beam pipe. The $x$-axis points from the IP to the
  centre of the LHC ring, and the $y$-axis points upward. Cylindrical
  coordinates $(R,\phi)$ are used in the transverse plane, $\phi$ being the
  azimuthal angle around the beam pipe. The pseudorapidity is defined in terms
  of the polar angle $\theta$ as $\eta=-\ln\tan(\theta/2)$.}%
  \bibAnnoteFile{Stop}{coordSys}%
\end{thebibliography}
\onecolumngrid
\clearpage 
\begin{flushleft}
{\Large The ATLAS Collaboration}

\bigskip

G.~Aad$^{\rm 48}$,
B.~Abbott$^{\rm 111}$,
J.~Abdallah$^{\rm 11}$,
S.~Abdel~Khalek$^{\rm 115}$,
A.A.~Abdelalim$^{\rm 49}$,
O.~Abdinov$^{\rm 10}$,
B.~Abi$^{\rm 112}$,
M.~Abolins$^{\rm 88}$,
O.S.~AbouZeid$^{\rm 158}$,
H.~Abramowicz$^{\rm 153}$
% ...
\bigskip

% ...
$^{10}$ Institute of Physics, Azerbaijan Academy of Sciences, Baku, Azerbaijan\\
$^{11}$ Institut de F\'isica d'Altes Energies and Departament de F\'isica de la Universitat Aut\`onoma  de Barcelona and ICREA, Barcelona, Spain\\
% ...
$^{48}$ Fakult\"{a}t f\"{u}r Mathematik und Physik, Albert-Ludwigs-Universit\"{a}t, Freiburg i.Br., Germany\\
$^{49}$ Section de Physique, Universit\'e de Gen\`eve, Geneva, Switzerland\\
% ...
$^{88}$ Department of Physics and Astronomy, Michigan State University, East Lansing MI, United States of America\\
% ...
$^{111}$ Homer L. Dodge Department of Physics and Astronomy, University of Oklahoma, Norman OK, United States of America\\
$^{112}$ Department of Physics, Oklahoma State University, Stillwater OK, United States of America\\
% ...
$^{115}$ LAL, Univ. Paris-Sud and CNRS/IN2P3, Orsay, France\\
% ...
$^{153}$ Raymond and Beverly Sackler School of Physics and Astronomy, Tel Aviv University, Tel Aviv, Israel\\
% ...
$^{158}$ Department of Physics, University of Toronto, Toronto ON, Canada\\
% ...
\end{flushleft}

\end{document}